\documentclass[runningheads]{llncs}

\usepackage{mathtools}
\usepackage{graphicx}
\usepackage[final,hidelinks]{hyperref}
\usepackage{multicol}
\usepackage{listings}
\usepackage{xcolor}
\usepackage{enumitem}
\usepackage{lstcoq}
\usepackage{proof-dashed}

\newcommand*{\rulebreak}{\\[1em]}
\newcommand{\eoe}{\hfill $\triangleleft$}

\definecolor{dkviolet}{rgb}{.5,0,.5}
\definecolor{dkblue}{rgb}{0,0,.5}
\definecolor{dkgreen}{rgb}{0,.5,0}
\definecolor{dkred}{rgb}{.5,0,0}
\definecolor{ltblue}{rgb}{0,.4,.6}

\def\orcidID#1{}

\begin{document}
\title{Certifying Choreography Compilation\thanks{Work partially supported by Villum Fonden, grant no.\ 29518.}}

\author{Luís Cruz-Filipe\orcidID{0000-0002-7866-7484}
\and
Fabrizio Montesi\orcidID{0000-0003-4666-901X}
\and
Marco Peressotti\orcidID{0000-0002-0243-0480}
}

\authorrunning{L. Cruz-Filipe et al.}

\institute{Department of Mathematics and Computer Science, University of Southern Denmark\\
Campusvej 55, 5230 Odense M, Denmark
\email{\{lcfilipe,fmontesi,peressotti\}@imada.sdu.dk}}

\maketitle

\begin{abstract}
Choreographic programming is a paradigm for developing concurrent and distributed systems, where programs are \emph{choreographies} that define, from a global viewpoint, the computations and interactions that communicating processes should enact.
\emph{Choreography compilation} translates choreographies into the local definitions of process behaviours, given as terms in a process calculus.

Proving choreography compilation correct is challenging and error-prone, because it requires relating languages in different paradigms (global interactions vs local actions) and dealing with a combinatorial explosion of proof cases.
We present the first certified program for choreography compilation for a nontrivial choreographic language supporting recursion.
\end{abstract}

\section{Introduction}\label{sec:intro}
Choreographic programming is an emerging programming paradigm where the desired communication behaviour of a system of communicating processes can be defined from a global viewpoint in programs known as \emph{choreographies}~\cite{M13p}.
Then, a provably-correct compiler can automatically generate executable code for each process, with the guarantee that executing these processes together implements the communications prescribed in the choreography~\cite{CHY12,CM13}.
The theory of such compilers is typically called EndPoint Projection (EPP).

Choreographies are inspired by the ``Alice and Bob'' notation for security protocols~\cite{NS78}.
The key idea is to have a linguistic primitive for a communication from a participant to another: statement \lstinline+Alice.e -> Bob.x+ reads ``\lstinline+Alice+ evaluates expression \lstinline+e+ and sends the result to \lstinline+Bob+, who stores it in variable \lstinline+x+''.
This syntax has two main advantages.
First, the desired communications are syntactically manifest in a choreography, which makes choreographic programming suitable for making interaction protocols precise.
Second, it disallows writing mismatched send and receive actions, so code generated from a choreography enjoys progress (the system never gets stuck)~\cite{CM13}.

The potential of choreographic programming has motivated the study of choreographic languages and EPP definitions for different applications, including self-adaptive systems~\cite{DGGLM17}, information flow~\cite{LN15}, system integration~\cite{GLR18}, parallel algorithms~\cite{CM16}, cyber-physical systems~\cite{GMP20,LH17,LNN16}, and security protocols~\cite{GMP20}.

EPP involves three elements: the source choreography language, the target language, and the compiler.
The interplay between these components, where a single instruction at the choreographic level might be implemented by multiple instructions in the target language, makes the theory of choreographic programming error-prone: for even simpler approaches, like abstract choreographies without computation, it has been recently discovered that a few key results published in peer-reviewed articles do not hold and their theories required adjustments~\cite{SY19}, raising concerns about the soundness of these methods.

In this article, we present a certified program for EPP, which translates terms of a Turing complete choreographic language into terms of a distributed process calculus.
Our main result is the formalisation of the hallmark result of choreographic programming, the ``EPP Theorem'': an operational correspondence between choreographies and their endpoint projections.
This is the first time that this result has been formalised in a theorem prover, increasing our confidence in the methodology of choreographies.

\paragraph{Structure and Related Work.}
Our formalisation is developed in Coq~\cite{CoqArt}, and we assume some familiarity with it.
We start from a previous formalisation~\cite{CMP21} of a choreographic language (Core Choreographies~\cite{CM20}), which we recap in Sect.~\ref{sec:background}.
This formalisation only deals with the choreographic language and its properties; in particular, the target calculus for EPP is not defined therein.

In Sect.~\ref{sec:sp}, we define our target language: a distributed process calculus inspired by the informal presentation in~\cite{M21itc}.
The calculus has communication primitives that recall those commonly used for implementing choreography languages, e.g., as found in (multiparty) session types~\cite{HYC16}.
In Sect.~\ref{sec:merging}, we define \emph{merging}~\cite{CHY12}: a partial operator that addresses the standard problem (for choreographies) of checking that each process implementation eventually agrees on the choice between alternative behaviours in protocols~\cite{BBO12,CHY12}.
Building on merging, in Sect.~\ref{sec:epp}, we define EPP.
Then, in Sect.~\ref{sec:pruning}, we explore \emph{pruning}~\cite{CHY12}: a preorder induced by merging that plays a key role in the EPP Theorem, proved in Sect.~\ref{sec:epp-theorem}.

Choreographies are used in industry for the specification and definition of web services and business processes~\cite{bpmn,wscdl}.
These languages feature recursion or loops, which are not present in the only other formalisation work on choreographies that we are aware of~\cite{GA18}.
We have validated the theory of EPP from~\cite{M21itc}, and made explicit several properties that are typically only implicitly assumed.
Our results show that the ideas developed by researchers on choreographies, like merging, can be relied upon for languages of practical appeal.

\section{Background}
\label{sec:background}

We use the choreographic language of~\cite{M21itc}, which is inspired by Core Choreographies (CC)~\cite{CM20}.
So far, this is the only Turing complete choreographic language that has been formalised~\cite{CMP21}.
In this section, we recap this formalisation, which our work builds upon.
We refer the reader to~\cite{CMP21} for a discussion of the design options both behind the choreographic language and the formalisation.
Some of these are relevant for our development, and we explain them when needed.

CC is designed to model communication protocols involving several participants, called processes, each equipped with memory cells, identified by variables.
Communications are of two kinds: \emph{value communications}, where a process evaluates an expression using the values stored in its memory and sends the result to a (distinct) process; and \emph{(label) selections}, where a process selects from different behaviours available at another process by means of an identifier (the label).
Selections are used to communicate local choices made by a process to other processes.
Recursive and infinite behaviour is achieved by defining procedures.

\paragraph{Syntax.}
The formalisation of CC is parametric on the types of processes (\lstinline!Pid!, ranged over by \lstinline!p!), variables (\lstinline!Var!, ranged over by \lstinline!x!), expressions (\lstinline!Expr!, ranged over by \lstinline!e!), values stored in memory (\lstinline!Value!, ranged over by \lstinline!v!), Boolean expressions (\lstinline!BExpr!, ranged over by \lstinline!b!) 
and procedure names (\lstinline!RecVar!, ranged over by \lstinline!X!).
Equality on these types must be decidable.
Labels, ranged over by \lstinline!l!, are either \lstinline!left! or \lstinline!right!, which is common in choreographies and session types~\cite{CLMSW16,CMS18,W14}.

The syntax of choreographies is defined by the following BNF grammar.
\begin{lstlisting}
eta ::= p#e --> q.x | p --> q[l]
  C ::= eta;; C | If p.b Then Ct Else Ce | Call X | RT_Call X ps C | End
\end{lstlisting}

An \lstinline!eta! is a communication action, where \lstinline+p#e --> q.x+ is a value communication\footnote{For readability, we use notations closer to the usual mathematical ones than in formalisation, where they are slightly different due to Coq's restrictions on overloading.} and \lstinline+p --> q[l]+ is a label selection.
\lstinline!Eta! is the type of all communication actions.

Choreographies are ranged over by \lstinline!C!.
A choreography \lstinline!eta;; C!, can execute a communication \lstinline!eta! and continue as \lstinline!C!.
A conditional \lstinline+If p.b Then Ct Else Ce+ evaluates the Boolean expression \lstinline+b+ in the memory of process \lstinline!p! and continues as \lstinline+Ct+ or \lstinline+Ce+, according to whether \lstinline+b+ evaluates to \lstinline+true+ or \lstinline+false+.
Choreography \lstinline+Call X+ is a call to a procedure \lstinline+X+.
Term \lstinline+RT_Call X ps C+ is a \emph{runtime term}, discussed below.
Term \lstinline+End+ is the terminated choreography.
We write \lstinline+Choreography+ for the type of all choreographies.

This grammar is implemented as a Coq inductive type, e.g., \lstinline+eta;; C+ stands for \lstinline+Interaction eta C+, where \lstinline+Interaction : Eta -> Choreography -> Choreography+ is a constructor of type \lstinline+Choreography+.

\begin{example}[Distributed Authentication]
\label{ex:authentication_chor}
The choreography \lstinline+C1+ below describes a multiparty authentication scenario where an identity provider \lstinline+ip+ authenticates a client \lstinline+c+, to server \lstinline+s+ (we name subterms for later use).%
\begin{lstlisting}
 C1 := c#credentials --> ip.x; If ip.(check x) Then C1t Else C1e
C1t := ip --> s[left]; ip --> c[left]; s#token --> c.t; End
C1e := ip --> s[right]; ip --> c[right]; End
\end{lstlisting}
\lstinline+C1+ starts with \lstinline+c+ communicating its credentials, stored in the local variable \lstinline+credentials+, to \lstinline+ip+, which stores them in \lstinline+x+.
Then, \lstinline+ip+ checks if the credentials are valid or not by evaluating the local expression \lstinline+check x+, %
and signals the result to \lstinline+s+ and \lstinline+c+ by selecting \lstinline+left+ when the credentials are valid (\lstinline+C1t+) and \lstinline+right+ otherwise (\lstinline+C1e+).
In the first case, the server communicates a token (stored in its local variable \lstinline+token+) to the client, %
otherwise the choreography ends.%

Because the guard of the conditional is evaluated by \lstinline+ip+, only this process knows which branch of the choreography to execute.
This is an instance of the \emph{knowledge of choice} problem.
Label selections propagate this information to processes whose behaviour depends on this choice (in this case, both \lstinline+s+ and \lstinline+c+).
\eoe
\end{example}

A \emph{program} is a pair \lstinline+(Procedures P,Main P):DefSet * Choreography+.
Elements of type \lstinline+DefSet+ (set of procedure definitions) map each \lstinline+X+ to a pair
containing the processes used in \lstinline+X+ (\lstinline+Vars P X+) and the choreography to be
executed (\lstinline+Procs P X+).
These procedure definitions can then be called from each other and from \lstinline+Main P+ (using
term~\lstinline+Call X+), allowing for the definition of recursive behaviour.

Executing procedure calls generates runtime terms.
Choreography \lstinline+Call X+ can reduce by a process \lstinline+p+ entering \lstinline+X+.
It becomes \lstinline+RT_Call X ps C+, where \lstinline+C+ is the definition of \lstinline+X+ and \lstinline+ps+ is the list of processes used in \lstinline+X+ (other than \lstinline+p+).
This term can then either reduce by another process entering the procedure (and being removed from \lstinline+ps+), or by executing some action in \lstinline+C+ that does not involve processes in \lstinline+ps+ (and \lstinline+C+ is updated).
When the last process in \lstinline+ps+ calls \lstinline+X+, \lstinline+RT_Call X ps C+ reduces to \lstinline+C+.
The runtime term \lstinline+RT_Call X ps C+ is not meant to be used when writing a choreography, and
in particular it should not occur in procedure definitions.
A choreography that does not include any such term is called \emph{initial}.

There are three kinds of restrictions when writing choreographies.
\begin{enumerate}[nosep,noitemsep,label={(\roman{*})}]
\item Intended use of choreographies.
  Interactions must have distinct processes (no self-communication), e.g., 
  \lstinline+p#e --> p.x+ is disallowed.

\item Intended use of runtime terms.
  All choreographies \lstinline+Procs P X+ must be initial.
  \lstinline+Main P+ may include subterms \lstinline+RT_Call X ps C+, but
  \lstinline+ps+ must be nonempty and include only process names that occur in
  \lstinline+Vars P X+.

\item Design choices in the formalisation.
  Informally, \lstinline+Vars X+ contains the processes that are used in \lstinline+Procs X+.
  This introduces constraints that are encapsulated in the definition of well-formed
  program.
\end{enumerate}
The constraints in the last category are particularly relevant for the proof of the EPP Theorem, so we discuss them in that context in Sect.~\ref{sec:epp-theorem}.

\begin{example}\label{ex:file_transfer_chor}
Let \lstinline!Defs:DefSet! map \lstinline!FileTransfer! to the pair consisting of the process list \lstinline!c :: s :: nil! and the following choreography. 
\begin{lstlisting}
s.(file, check) --> c.x;                    (* send file and check data          *)
If c.(crc(fst(x)) == snd(x))              (* cyclic redundancy check           *)
  Then c --> s[left]; End                   (* file received correctly, end      *)
  Else c --> s[right]; Call FileTransfer    (* errors detected, retry            *)
\end{lstlisting}
\lstinline+FileTransfer+ describes a file transfer protocol between a server \lstinline+s+ and a client \lstinline+c+ using Cyclic Redundancy Checks (\lstinline+crc+) to detect errors from a noisy channel.
\eoe
\end{example}

\paragraph{Semantics.}
The semantics of CC is defined as a labelled transition system using inductive types.
It uses a \emph{state}, which is a function mapping process variables to their values: \lstinline+State:Pid -> Var -> Value+.

The semantics is structured in three layers.
The first layer specifies transitions with the following relation, parameterised on a
set of procedure definitions.
\begin{lstlisting}
CCC_To : DefSet -> Choreography -> State
				-> RichLabel -> Choreography -> State -> Prop
\end{lstlisting}
Rules in this set include that \lstinline+p#e --> q.x;; C+ in state \lstinline+s+
transitions to \lstinline+C+ in state \lstinline+s'+, where \lstinline+s'+ coincides with \lstinline+s+
except that \lstinline+s' q x+ now stores the value obtained by evaluating \lstinline+e+
at \lstinline+p+ in state \lstinline+s+.\footnote{The semantics of CC only requires extensional
  equality of states, which is why \lstinline+s'+ is quantified over rather than directly defined
  from \lstinline+s+.}
\lstinline+RichLabel+ includes information about the executed term -- the above communication is labelled \lstinline+R_Com p v q x+.
There are also rules for out-of-order execution: interactions involving distinct processes can be executed in any order, reflecting concurrency.
For example, \lstinline+p#e --> q.x;; r --> s[left]+ can execute as a communication from \lstinline+p+ to \lstinline+q+ followed by a label selection from \lstinline+r+ to \lstinline+s+, but also as the latter label selection followed by the former communication.

The second layer raises transitions to the level of programs, abstracting from unobservable details.
It is defined by a single rule: if \lstinline+CCC_To Defs C s t C' s'+, then \lstinline+({|Defs; C|},s) --[forget t]--> ({|Defs; C'|},s')+.
Here, \lstinline+{|Defs; C|}+ denotes the program built from \lstinline+Defs+ and \lstinline+C+, and \lstinline+forget+ removes unobservable details from transition labels (e.g., \lstinline+forget (R_Com p v q x)+ is \lstinline+L_Com p v q+).
Labels for conditionals and procedure calls all simplify to \lstinline+L_Tau p+, denoting an internal action at \lstinline+p+.
The third layer defines the reflexive and transitive closure of program transitions.

Important properties of CC formalised in~\cite{CMP21} include deadlock-freedom by design (any program \lstinline+P+ such that \lstinline+Main P <> End+ reduces), confluence (if \lstinline+P+ can execute two different sequences of actions, then the resulting programs can always reduce to a common program and state), and Turing completeness.

\begin{example}
  \label{ex:auth_chor_red}
Consider the program \lstinline+{|D; C1|}+ where \lstinline+C1+ is the choreography in Example~\ref{ex:authentication_chor} and \lstinline+D:DefSet+ is arbitrary (there are no recursive calls in \lstinline+C1+).
\begin{lstlisting}
{|D; C1|},st1) --[L_Com c ip v1]--> ({|D; If ip.(check x) Then C1t Else C1e|},st2)
\end{lstlisting}
where \lstinline+v1+ is the evaluation of \lstinline+e+ at \lstinline+c+ in \lstinline+st1+ and \lstinline+st2+ updates the value of \lstinline+ip+'s variable \lstinline+x+ accordingly.
If \lstinline+check x+ is \lstinline+true+ at \lstinline+ip+ in \lstinline+st2+, then it continues as follows.
\begin{lstlisting}
{|D; If ip.(check x) Then C1t Else C1e|}, st2) --[L_Tau ip]--> {|D; C1t|}, st2)
  --[L_Sel ip s left; L_Sel ip c left]-->** {|D; s#token --> c.t; End|}, st2)
  --[L_Com s c v2]--> {|D; End|}, st3)
\end{lstlisting}
where \lstinline+v2+ is the evaluation of \lstinline+token+ at \lstinline+s+ in \lstinline+st2+ and \lstinline+st3+ updates \lstinline+st2+ accordingly.
Otherwise, it continues as follows.
\begin{lstlisting}
{|D; If ip.(check x) Then C1t Else C1e|}, st2) --[L_Tau ip]--> {|D; C1e|}, st2)
  --[L_Sel ip s right; L_Sel ip c right]-->** {|D; End|}, st2)
\end{lstlisting}
In compound transitions, the actions in the label are executed in order.
\eoe
\end{example}

\begin{example}
Let \lstinline+Defs+ as in Example~\ref{ex:file_transfer_chor} and \lstinline+C+ be the body of \lstinline+FileTransfer+. 
Consider the program \lstinline+{|Defs; Call FileTransfer|}+.
The processes in the procedure \lstinline+FileTransfer+ can start a call in any order as exemplified by the transitions below.
\begin{lstlisting}
{|Defs; Call FileTransfer|},st) --[L_Tau c]-->
  ({|Defs; RT_Call FileTransfer s::nil C|},st) --[L_Tau s]--> ({|Defs; C|},st)
\end{lstlisting}
\begin{lstlisting}
{|Defs; Call FileTransfer|},st) --[L_Tau s]-->
  ({|Defs; RT_Call FileTransfer c::nil C|},st) --[L_Tau c]--> ({|Defs; C|},st)
\end{lstlisting}
The state \lstinline+st+ is immaterial.\eoe
\end{example}

\section{Stateful Processes}
\label{sec:sp}

Our first contribution is formalising SP (Stateful Processes), the process calculus for implementing CC~\cite{M21itc}, which is inspired by the process calculi in~\cite{CM17,CM20}.\footnote{The choreography language CC is also inspired by the works~\cite{CM17,CM20}, but formalising it in Coq benefitted substantially from adopting a labelled transition system semantics. This is discussed extensively in~\cite{CMP21}. In this work, we made similar changes to the process calculus not only for similar reasons, but also to keep a close correspondence with the choreography language.}

\subsection{Syntax}
SP is defined as a Coq functor with the same parameters as CC.
Its syntax contains three ingredients: behaviours, defining the actions performed by individual processes; networks, consisting of several processes running in parallel; and programs, defining a set of procedure definitions that all processes can use.

\paragraph{Behaviours.}
Behaviours are sequences of local actions -- counterparts to the terms that can be written in CC --
defined by the following BNF grammar.
\begin{lstlisting}
B ::= End | p!e;B | p?x;B | p(+)l;B | p&mBl//mBr | If b Then Bt Else Be | Call X
\end{lstlisting}
This grammar is again formalised as a Coq inductive type, called \lstinline+Behaviour+.
Terms \lstinline+End+, \lstinline+If b Then Bt Else Be+ and \lstinline+Call X+ are as in CC.
Value communications are split into \lstinline+p!e; B+, which evaluates \lstinline+e+, sends the
result to \lstinline+p+, and continues as \lstinline+B+, and by \lstinline+p?x; B+, which receives a
value from \lstinline+p+, stores it at \lstinline+x+, and continues as
\lstinline+B+.\footnote{Processes communicate by name. In practice, names can be either process
  identifiers (cf.\ actors), network addresses, or session correlation data.}

Label selections are divided into \lstinline!p(+)l; B! -- sending the label \lstinline+l+ to
\lstinline+p+ and continuing as \lstinline+B+ -- and \lstinline+p & mBl // mBr+, where one of
\lstinline+mBl+ or \lstinline+mBr+ is chosen according to the label selected by \lstinline+p+.
Both \lstinline+mBl+ and \lstinline+mBr+ have type \lstinline+option Behaviour+ (either
\lstinline+None+ or \lstinline+Some B+, where \lstinline+B+ is a behaviour): a process does not
need to offer behaviours corresponding to all possible labels.
Informally, branching terms are partial functions from labels to behaviours; we capitalise on the
fact that there are only two labels to simplify the formalisation.

Many results about \lstinline+Behaviour+s are proved by structural induction, requiring inspection of subterms of branching terms.
The induction principle automatically generated by Coq is not strong enough for this, and our formalisation includes a stronger result that we use in later proofs.

\paragraph{Networks.}
Networks have type \lstinline+Pid -> Behaviour+.
We define extensional equality of networks, \lstinline+N == N'+, and show that it is an equivalence relation.
To model practice, where networks are written as finite parallel compositions of behaviours, we introduce some constructions: \lstinline+N|N'+ is the parallel composition of \lstinline+N+ and \lstinline+N'+; \lstinline+p[B]+ is the network mapping \lstinline+p+ to \lstinline+B+ and all other processes to \lstinline+End+; and \lstinline+N ~~ p+ denotes network \lstinline+N+ where \lstinline+p+ is now mapped to \lstinline+End+.
The formalisation includes a number of lemmas about extensional equality, for example that updating
the behaviours of two processes yields the same result independent of the order of the updates.

\paragraph{Programs.}
Finally, a \lstinline+Program+ is a pair \lstinline+(Procs P,Net P):DefSetB * Network+, where \lstinline+DefSetB = RecVar -> Behaviour+ maps procedure names to \lstinline+Behaviour+s.

Programs, networks and behaviours should satisfy well-formedness properties similar to those of CC.
However, these properties are automatically ensured when networks are automatically generated from
choreographies, so we do not discuss them here.
They are included in the formalisation.

\begin{example}\label{ex:authentication_net}
Consider the network \lstinline+N = c[Bc] | s[Bs] | ip[Bip]+ composed by the behaviours below.
\begin{lstlisting}
 Bc := ip!credentials; ip & Some (s?t; End) // Some End
 Bs := ip & Some (c!token; End) // Some End

Bip := c?x; Bip'
Bip' := If (check x) Then (s(+)left; c(+)left; End) Else (s(+)right; c(+)right; End)
\end{lstlisting}
This network implements the choreography in Example~\ref{ex:authentication_chor}.
\eoe
\end{example}

\subsection{Semantics}

The semantics of SP is defined by a labelled transition system.
Transitions for communications match dual actions in two processes, while conditionals and procedure
calls simply run locally.
We report the representative cases -- the complete definition can be found in the source
formalisation~\cite{CMP21-source}.
For readability, we first present them in the standard rule notation.

\begin{gather*}
\infer[\mbox{\lstinline+S_Com+}]{
	\mbox{\lstinline+(\{|Defs; N|\}, s) --[L_Com p v q]--> (\{|Defs; N'|\}, s')+}
}{
	\begin{array}{c}
	\mbox{\lstinline+v := eval_on_state e s p+}
	\quad
    \mbox{\lstinline+N p = q!e; B+}
    \quad
    \mbox{\lstinline+N q = p?x; B'+}
    \\
    \mbox{\lstinline+N' == N ~~ p ~~ q | p[B] | q[B']+}
    \quad
    \mbox{\lstinline+s' == update s q x v+}
    \end{array}
}
\rulebreak
\infer[\mbox{\lstinline+S_LSel+}]{
	\mbox{\lstinline+(\{|Defs; N|\}, s) --[L_Sel p q left]--> (\{|Defs; N'|\}, s')+}
}{
	\begin{array}{c}
	\mbox{\lstinline!N p = q(+)left; B!}
    \quad
    \mbox{\lstinline+N q = p & Some Bl // Br+}
    \\
    \mbox{\lstinline+N' == N ~~ p ~~ q | p[B] | q[Bl]+}
    \quad
    \mbox{\lstinline+s == s'+}
    \end{array}
}
\rulebreak
\infer[\mbox{\lstinline+S_Then+}]{
	\mbox{\lstinline+(\{|Defs; N|\}, s) --[L_Tau p]--> (\{|Defs; N'|\}, s')+}
}{
	\begin{array}{c}
	\mbox{\lstinline+N p = If b Then B1 Else B2+}
	\quad
	\mbox{\lstinline+beval_on_state b s p = true+}
	\\
    \mbox{\lstinline+N' == N ~~ p | p[B1]+}
    \quad
    \mbox{\lstinline+s == s'+}
    \end{array}
}
\end{gather*}
\begin{gather*}
\rulebreak
\infer[\mbox{\lstinline+S_Call+}]{
	\mbox{\lstinline+(\{|Defs; N|\}, s) --[L_Tau X p]--> (\{|Defs; N'|\}, s')+}
}{
	\mbox{\lstinline+N p = Call X+}
	&
        \mbox{\lstinline+N' == N ~~ p | p[Defs X]+}
        &
        \mbox{\lstinline+s == s'+}
}
\end{gather*}

Functions \lstinline+eval_on_state+ and \lstinline+beval_on_state+ evaluate a (Boolean) expression locally, given a state, while \lstinline+update s q x v+ updates state \lstinline+s+ by changing the value of \lstinline+q+'s variable \lstinline+x+ to \lstinline+v+.
We write also \lstinline+ ==+ for extensional equality of states.

As for choreographies, the formalisation of these rules is done in two steps.
The first step defines a transition relation parameterised on the set of procedure definitions, and
includes richer transition labels (necessary for doing case analysis on transitions).
This relation is defined as an inductive type, whose first defining clause is shown below.\footnote{Coq's type inference mechanism allows us to omit types of most universally quantified variables and parameters. We abuse this possibility to lighten the presentation.}

\begin{lstlisting}
Inductive SP_To (Defs : DefSetB) :
  Network -> State -> RichLabel -> Network -> State -> Prop :=
  | S_Com N p e B q x B' N' s s' : let v := (eval_on_state e s p) in
    N p = (q!e; B) -> N q = (p?x; B') ->  N' == N ~~ p ~~ q | p[B] | q[B'] ->
    s' == (update s q x v) -> SP_To Defs N s (R_Com p v q x) N' s'    (...)
\end{lstlisting}
This relation is then lifted to configurations just as for CC: if
\lstinline+SP_To Defs N s t N' s'+, then
\lstinline+({|Defs; N|},s) --[forget t]--> ({|Defs; N'|},s')+.
Closure under reflexivity and transitivity (with similar notation) is again defined as for
choreographies.

\begin{example}
  We illustrate the possible transitions of the network from Example~\ref{ex:authentication_net}.
  We abbreviate the behaviours of processes that do not change in a reduction to \lstinline+...+ to make it more clear what parts of the network are changed.
  Furthermore, we omit trailing \lstinline+End+s in \lstinline+Behaviour+s.

  The network starts by performing the transition
\begin{lstlisting}
{|D; c[Bc] | s[Bs] | ip[Bip]|}, st1) --[L_Com c ip v1]-->
  {|D; c[ip & Some (s?t) // Some End] | s[...] | ip[Bip']|}, st2)
\end{lstlisting}
where \lstinline+v1+ and \lstinline+st2+ are as in Example~\ref{ex:auth_chor_red}.
If \lstinline+eval_on_state (check x) st2 ip+ is \lstinline+true+, it continues as follows
\begin{lstlisting}
{|D; c[ip & Some (s?t) // Some End] | s[Bs] | ip[Bip'], st2) 
  --[L_Tau ip]-->             {|D; c[...] | s[...] | ip[s(+)left;c(+)left], st2)
  --[L_Sel ip s left]--> {|D; c[...] | s[c!token] | ip[c(+)left], st2)
  --[L_Sel ip c left]--> {|D; c[s?t] | s[...] | ip[End], st2)
  --[L_Com s c v2]-->         {|D; c[End] | s[End] | ip[End], st3)
\end{lstlisting}
where \lstinline+v2+ and \lstinline+st3+ are again as in Example~\ref{ex:auth_chor_red}.
Otherwise, it continues as follows.
\begin{lstlisting}
{|D; c[ip & Some (s?t) // Some End] | s[Bs] | ip[Bip'], st2) 
  --[L_Tau ip]-->              {|D; c[...] | s[...] | ip[s(+)right;c(+)right], st2)
  --[L_Sel ip s right]--> {|D; c[...] | s[End] | ip[c(+)right], st2)
  --[L_Sel ip c right]--> {|D; c[End] | s[End] | ip[End], st2)
\end{lstlisting}
The labels in these reductions are exactly as in Example~\ref{ex:auth_chor_red}.
\eoe
\end{example}

Transitions are compatible with network equality and state equivalence.
\begin{lstlisting}
Lemma SPP_To_eq : forall P s1 tl P' s2 s1' s2',
  s1 == s1' -> s2 == s2' -> (P,s1) --[tl]--> (P',s2) -> (P,s1') --[tl]--> (P',s2').

Lemma SPP_To_Network_eq : forall P1 P1' P2 s s' tl, (Net P1 == Net P1') ->
  (Procs P1 = Procs P1') -> (P1,s) --[tl]--> (P2,s') -> (P1',s) --[tl]--> (P2,s').
\end{lstlisting}
These results are instrumental in some of the later proofs, and are proven for the three levels of
reductions.
The following, related, result is also important.
\begin{lstlisting}
Lemma SPP_To_Defs_stable : forall Defs' N N' tl s s',
  {|Defs N,s|} --[tl]--> {|Defs' N',s'|} -> Defs = Defs'.
\end{lstlisting}

We also prove that transitions are completely determined by the label.
\begin{lstlisting}
Lemma SPP_To_deterministic : forall P s tl P' s' P'' s'',
  (P,s) --[tl]--> (P',s') -> (P,s) --[tl]--> (P'',s'') ->
  (Net P' == Net P'') /\ Procs P' = Procs P'' /\ (s' == s'').
\end{lstlisting}
Finally, we show that the semantics of SP is confluent.
Although this is not required for our main theorem, it is a nice result that confirms our
expectations.

The formalisation of SP consists of 37 definitions, 80 lemmas, and 2000 lines.

\section{Merging}
\label{sec:merging}

Intuitively, process implementations are generated from choreographies recursively, by projecting
each action in the choreography to the corresponding process action -- for example, a value
communication \lstinline+p#e --> q.x+ should be projected as a send action \lstinline+q!e+ at
\lstinline+p+, as a receive action \lstinline+p?x+ at \lstinline+q+, and not projected to any other
processes.
However, this causes a problem with conditionals.
Projecting a choreography \lstinline+If p.b Then Ct Else Ce+ for any process other than
\lstinline+p+, say \lstinline+q+, requires combining the projections obtained for \lstinline+Ct+ and \lstinline+Ce+, such that \lstinline+q+ can ``react'' to whichever choice \lstinline+p+ will make.
This combination is called \emph{merging}~\cite{CHY12}.

Merge is typically defined as a partial function mapping pairs of behaviours to behaviours that
returns a behaviour combining all possible executions of the two input behaviours (if possible).
For SP, two behaviours can be merged if they are structurally similar, with the possible exception of branching terms: we can merge branchings that offer options on distinct labels.
For example, merging \lstinline+p & (Some B) // None+ with \lstinline+p & None // (Some B')+ yields \lstinline+p & (Some B) // (Some B')+,
allowing \lstinline+If p??b Then (p-->q[left];; q.e --> p;; End) Else (p-->q[right];; End)+ to be projected for \lstinline+q+ as \lstinline+p & Some (p!e; End) // Some End+.

Since functions in Coq are total, formalising merging poses a problem.
Furthermore, assigning type \lstinline+Behaviour -> Behaviour -> option Behaviour+ to merging causes ambiguity, since branching terms have subterms of type \lstinline+option Behaviour+.\footnote{Essentially because it is not possible to distinguish if the behaviour assigned to a label is \lstinline+None+ because it was not defined, or because a recursive call to merge failed.}
Instead, we define a type \lstinline+XBehaviour+ of extended behaviours, defined as \lstinline+Behaviour+ with an extra constructor \lstinline+XUndefined : XBehaviour+.
Thus, an \lstinline+XBehaviour+ is a \lstinline+Behaviour+ that may contain \lstinline+XUndefined+
subterms.

The connection between \lstinline+Behaviour+ and \lstinline+XBehaviour+ is established by means of
two functions: \lstinline+inject : Behaviour -> XBehaviour+, which isomorphically injects each
constructor of \lstinline+Behaviour+ into the corresponding one in \lstinline+XBehaviour+, and
\lstinline+collapse : XBehaviour -> XBehaviour+ that maps all \lstinline+XBehaviour+s with
\lstinline+XUndefined+ as a subterm to \lstinline+XUndefined+.
The most relevant properties of these functions are:
\begin{lstlisting}
Lemma inject_elim : forall B, exists B', inject B = B' /\ B' <> XUndefined.
Lemma collapse_inject : forall B, collapse (inject B) = inject B.
Lemma collapse_char'' : forall B, collapse B = XUndefined -> forall B', B <> inject B'.
Lemma collapse_exists : forall B, collapse B <> XUndefined -> exists B', B = inject B'.
\end{lstlisting}

Using this type, we first define \lstinline+XMerge+ on \lstinline+XBehaviour+s as below, where
we report only representative cases.
(\lstinline+Pid_dec+ and \lstinline+Expr_dec+ are lemmas stating decidability of equality on
\lstinline+Pid+ and \lstinline+Expr+, allowing us to do case analysis.)
\begin{lstlisting}
Fixpoint Xmerge (B1 B2:XBehaviour) : XBehaviour := match B1, B2 with
| XEnd, XEnd => XEnd
| XSend p e B, XSend p' e' B' => if Pid_dec p p' && Expr_dec e e'
    then match Xmerge B B' with XUndefined => XUndefined
                              | _ => XSend p e (Xmerge B B') end
    else XUndefined
| XBranching p Bl Br, XBranching p' Bl' Br' => if Pid_dec p p'
    then let BL := match Bl with None => Bl' | Some B =>
        match Bl' with None => Bl | Some B' => Some (Xmerge B B') end end
      in let BR := match Br with None => Br' | Some B =>
        match Br' with None => Br | Some B' => Some (Xmerge B B') end end
      in match BL, BR with Some XUndefined, _ => XUndefined
         | _, Some XUndefined => XUndefined | _, _ => XBranching p BL BR end
    else XUndefined                                                        (...)
\end{lstlisting}
Using \lstinline+XMerge+ we can straightforwardly define merging.
\begin{lstlisting}
Definition merge B1 B2 := Xmerge (inject B1) (inject B2).
\end{lstlisting}

We show \lstinline+merge+ to be idempotent, commutative and associative.
This last proof illustrates the major challenge of this stage of the formalisation: it requires a triple induction with $512$ cases, of which $84$ cannot be solved automatically.
These had to be divided in further subcases of different levels of complexity.
The final case, when all behaviours are branching terms, requires six (!) nested inductions to generate the $64$ possible combinations of defined/undefined branches, which had to be done by hand.
The total proof is over 500 lines.

The largest set of lemmas about \lstinline+merge+ deals with inversion
results, such as:
\begin{lstlisting}
Lemma merge_inv_Send : forall B B' p e X, merge B B' = XSend p e X ->
  exists B1 B1', B = p ! e; B1 /\ B' = p ! e; B1' /\ merge B1 B1' = X.
\end{lstlisting}
The similar result for branching terms is much more complex, since there are several cases for each branch (if it is \lstinline+None+, then both \lstinline+B+ and \lstinline+B'+ must have \lstinline+None+ in the corresponding branch, otherwise it may be from \lstinline+B+, from \lstinline+B'+, or a merge of both), and its proof suffers from the same problems as the proof of associativity (thankfully, not to such a dramatic level).
Automation works better here, and the effect of the large number of subcases is mostly felt in the time required by the \lstinline+auto+ tactic.
Still, the formalisation of merging consists of 6 definitions, 43 lemmas, and 2550 lines %
-- giving an average proof length of over 50 lines.

\section{EndPoint Projection}
\label{sec:epp}

The next step is defining EndPoint Projection (EPP): a partial function that maps programs in CC to programs in SP. %
The target instance of SP has the same parameters as CC, except that the set of procedure names is \lstinline+RecVar * Pid+ -- each procedure is implemented from the point of view of each process in it.

Partiality of EPP stems from the problem of choreography realisability~\cite{BBO12}.
A choreography such as \lstinline+If p.b Then (q#e --> r.x) Else End+ cannot be implemented without additional communications between \lstinline+p+, \lstinline+q+ and \lstinline+r+, since the latter processes need to know the result of the conditional to decide whether to communicate (see also Example~\ref{ex:authentication_chor}).
We say that this choreography is not \emph{projectable}~\cite{CM20}.

We define EPP in several layers.
First, we define a function \lstinline+bproj : DefSet ->+ \lstinline+Choreography -> Pid -> XBehaviour+ projecting the behaviour of a single process.
Intuitively, \lstinline+bproj Defs C p+ attempts to construct \lstinline+p+'s behaviour as specified by \lstinline+C+; the parameter \lstinline+Defs+ is used for procedure calls, whose projections depend on whether \lstinline+p+ participates in the procedure.
Returning an \lstinline+XBehaviour+ instead of an \lstinline+option Behaviour+ gives information about where exactly merging fails (the location of \lstinline+XUndefined+ subterms), which can be used for debugging, providing information to programmers, or for automatic repair of choreographies~\cite{BB16,CM20}.

We show some illustrative cases of the definition of \lstinline+bproj+.
\begin{lstlisting}
Fixpoint bproj (Defs:DefSet) (C:Choreography) (r:Pid) : XBehaviour := match C with
| p#e --> q.x;; C' =>
  if Pid_dec p r then XSend q e (bproj Defs C' r)
                else if Pid_dec q r then XRecv p x (bproj Defs C' r)
                                     else bproj Defs C' r
| p --> q[left];; C' =>
  if Pid_dec p r then XSel q left (bproj Defs C' r)
                 else if Pid_dec q r then XBranching p (Some (bproj Defs C' r)) None
                                      else bproj Defs C' r
| If p.b Then C1 Else C2 =>
  if Pid_dec p r then XCond b (bproj Defs C1 r) (bproj Defs C2 r)
                 else Xmerge (bproj Defs C1 r) (bproj Defs C2 r)
| CCBase.Call X => if In_dec P.eq_dec r (fst (Defs X)) then XCall (X,r) else XEnd
(...)
\end{lstlisting}

The next step is generating projections for all relevant processes.
We take the set of processes as a parameter, and collapse all individual projections.
\begin{lstlisting}
Definition epp_list (Defs:DefSet) (C:Choreography) (ps:list Pid)
	: list (Pid * XBehaviour) := map (fun p => (p, collapse (bproj Defs C p))) ps.
\end{lstlisting}

A choreography \lstinline+C+ is projectable wrt \lstinline+Defs+ and \lstinline+ps+ if \lstinline+epp_list Defs C ps+ does not contain \lstinline+XUndefined+, and \lstinline+Defs:DefSet+ is projectable wrt a set of procedure names \lstinline+Xs+ if \lstinline+snd (Defs X)+ is projectable wrt \lstinline+Defs+ and \lstinline+fst (Defs X)+ for each \lstinline+X+ in \lstinline+Xs+.
Projectability of programs is a bit more involved, and we present its Coq formalisation before discussing it.
\begin{lstlisting}
Definition projectable Xs ps P :=
  projectable_C (Procedures P) ps (Main P) /\ projectable_D Xs (Procedures P) /\
  (forall p, In p (CCC_pn (Main P) (fun _ => nil)) -> In p ps) /\
  (forall p X, In X Xs -> In p (fst (Procedures P X)) -> In p ps) /\
  (forall p X, In X Xs -> In p (CCC_pn (snd (Procedures P X)) (fun _ => nil)) -> In p ps).
\end{lstlisting}
The first two conditions simply state that \lstinline+Main P+ and \lstinline+Procedures P+ are
projectable wrt the appropriate parameters.
The remaining conditions state that the sets \lstinline+ps+ and \lstinline+Xs+ include all processes
used in \lstinline+P+ and all procedures needed to execute \lstinline+P+.
(These sets are not necessarily computable, since \lstinline+Xs+ is not required to be finite.
However, in practice, these parameters are known -- so it is easier to include them in the definition.)
Function \lstinline+CCC_pn+ returs the set of processes used in a choreography, given the sets of
processes each procedure call is supposed to use.

We now define compilation of projectable choreographies (\lstinline+epp_C+), projectable sets of procedure definitions (\lstinline+epp_D+), and projectable programs (\lstinline+epp+).
These definitions depend on proof terms whose structure needs to be explored, and are done interactively; afterwards, we show them to be independent of the proof terms, and that they work as expected.
We give a few examples.

\begin{lstlisting}
Lemma epp_C_wd : forall Defs C ps H H', (epp_C Defs ps C H) == (epp_C Defs ps C H').
Lemma epp_C_Com_p : forall Defs ps C p e q x HC HC', In p ps ->
  epp_C Defs ps (p#e-->q.x;;C) HC p = q!e; epp_C Defs ps C HC' p.
Lemma epp_C_Cond_r : forall Defs ps p b C1 C2 HC HC1 HC2 r, p <> r ->
  inject (epp_C Defs ps (If p.b Then C1 Else C2) HC r)
  = merge (epp_C Defs ps C1 HC1 r) (epp_C Defs ps C2 HC2 r).
\end{lstlisting}

Projectability of \lstinline+C+ does not imply projectability of choreographies that \lstinline+C+ can transition to.
This is due to the way runtime terms are projected: \lstinline+RT_Call X ps C'+ is projected as a call to \lstinline+(X,p)+ if \lstinline+p+ is in \lstinline+ps+, and as the projection of \lstinline+C'+ otherwise.
Our definition of projectability allows in principle for \lstinline+C+ to be unprojectable for a process in \lstinline+ps+, which would make it unprojectable after transition.
That this situation does not arise is a consequence of the intended usage of runtime terms: initially \lstinline+C'+ is obtained from the definition of a procedure, and \lstinline+ps+ is the set of processes used in this procedure.
Afterwards \lstinline+ps+ only shrinks, while \lstinline+C'+ may change due to execution of actions outside \lstinline+ps+.
We capture these conditions in the notion of strong projectability, whose representative case is:
\begin{lstlisting}
Fixpoint str_projectable Defs (C:Choreography) (r:Pid) : Prop := 
match C with | RT_Call X ps C => str_projectable Defs C r /\
     (forall p, In p ps -> In p (fst (Defs X))
          /\ Xmore_branches (bproj Defs (snd (Defs X)) p) (bproj Defs C p))      (...)
\end{lstlisting}
The relation \lstinline+Xmore_branches+, explained in the next section, is a semantic characterisation of how the projection of \lstinline+bproj Defs snd (Defs X) p+ may change due to execution of actions not involving \lstinline+p+ in \lstinline+snd (Defs X)+.

Projectability and strong projectability coincide for initial choreographies.
Furthermore, we state and prove lemmas that show that strong projectability of \lstinline+C+ imply
strong projectability of any choreography that \lstinline+C+ can transition to.

\section{Pruning}
\label{sec:pruning}

The key ingredient for our correspondence result is a relation on behaviours usually called \emph{pruning} \cite{CHY12,CM13}.
Pruning relates two behaviours that differ only in that one offers more options in branching terms
than the other; we formalise this relation with the more suggestive name of
\lstinline+more_branches+ (in line with~\cite{M21itc}), and we include some illustrative cases of its definition.
\begin{lstlisting}
Inductive more_branches : Behaviour -> Behaviour -> Prop :=
| MB_Send p e B B': more_branches B B' -> more_branches (p ! e; B) (p ! e; B')
| MB_Branching_None_None p mBl mBr :
    more_branches (p & mBl // mBr) (p & None // None)
| MB_Branching_Some_Some p Bl Bl' Br Br' :
    more_branches Bl Bl' -> more_branches Br Br' ->
    more_branches (p & Some Bl // Some Br) (p & Some Bl' // Some Br')           (...)
\end{lstlisting}
The need for pruning arises naturally when one considers what happens when a choreography \lstinline+C+ executes a conditional at a process \lstinline+p+.
In the continuation, only one of the branches is kept.
However, no process other than \lstinline+p+ knows that this action has been executed; therefore, if the projection of \lstinline+C+ executes the corresponding action, both behaviours are still available for all processes other than~\lstinline+p+.
Given how merging is defined, this means that these processes' behaviours may contain more branches than those of the projection of the choreography after reduction.

Pruning is naturally related to merging, as stated in the following lemmas.
\begin{lstlisting}
Lemma more_branches_char : forall B B', more_branches B B' <-> merge B B' = inject B.
Lemma merge_more_branches :
  forall B1 B2 B, merge B1 B2 = inject B -> more_branches B B1.
\end{lstlisting}

Pruning is also reflexive and transitive.
Further, if two behaviours have an upper bound according to pruning, then their merge is defined and is their lub.
\begin{lstlisting}
Lemma more_branches_merge : forall B B1 B2, more_branches B B1 ->
  more_branches B B2 -> exists B', merge B1 B2 = inject B' /\ more_branches B B'.
\end{lstlisting}

Finally, two behaviours with fewer branches than two mergeable behaviours are themselves mergeable.
\begin{lstlisting}
Lemma more_branches_merge_extend : forall B1 B2 B1' B2' B,
  more_branches B1 B1' -> more_branches B2 B2' -> merge B1 B2 = inject B ->
  exists B', merge B1' B2' = inject B' /\ more_branches B B'.
\end{lstlisting}

These two results are key ingredients to the cases dealing with conditionals in the proof of the EPP
Theorem.
They require extensive case analysis ($512$ cases for the last lemma, of which $81$ are not
automatically solved by Coq's \lstinline+inversion+ tactic even though most are contradictory).
Analogous versions of some of these lemmas also need to be extended to \lstinline+XBehaviour+s,
which is straightforward.

Pruning extends pointwise to networks, which we denote as \lstinline+N >> N'+.
The key result is that, due to how the semantics of SP is defined, pruning a network cannot add new
transitions.
\begin{lstlisting}
Lemma SP_To_more_branches_N : forall Defs N1 s N2 s' Defs' N1' tl,
  SP_To Defs N1 s tl N2 s' -> N1' >> N1 -> (forall X, Defs X = Defs' X) ->
  exists N2', SP_To Defs' N1' s tl N2' s' /\ N2' >> N2.
\end{lstlisting}
The reciprocal of this result only holds for choreography projections, and is proven after the
definition of EPP.

The formalisation of pruning includes 3 definitions, 25 lemmas, and 950 lines.
Some of these results are used for defining of EPP (previous section), but we delayed their presentation as its motivation is clearer after seeing those definitions.

\begin{example}
The choreography \lstinline+C1+ in Example~\ref{ex:authentication_chor} is projectable, yielding the network in Example~\ref{ex:authentication_net}:
\lstinline+bproj Defs C1 c = inject Bc+, \lstinline+bproj Defs C1 s = inject Bs+, and \lstinline+bproj Defs C1 ip = inject Bip+.

If we remove selections from \lstinline+C1+, the resulting choreography is not projectable on process \lstinline+c+: projecting the conditional requires merging the projections at \lstinline+c+ of the two branches (now simply \lstinline+s.token --> c.t; End+ and \lstinline+End+), which fails since \lstinline+bproj Defs C5 c = inject (s?t; End)+ and \lstinline+bproj Defs End c = inject End+, but \lstinline+merge (s?t; End) End = XUndefined+.
Likewise, merging would fail for \lstinline+s+.
\eoe
\end{example}

\section{EPP Theorem}
\label{sec:epp-theorem}

We now prove the operational correspondence between choreographies and their projections, in two directions: if a choreography can make a transition, then its projection can make the same transition; and if the
projection of a choreography can make a transition, then so can the choreography.
The results of the transitions are not directly related by projection, since choreography transitions
may eliminate some branches in the projection; thus, establishing the correspondence for multi-step transitions
requires some additional lemmas on pruning.

\paragraph{Preliminaries.}
Both directions of the correspondence depend on a number of results relating choreography
transitions and their projections.
These results follow a pattern: the results for communications state a precise
correspondence; the ones for conditionals include pruning in their conclusions; and the ones for
procedure calls require additional hypotheses on the set of procedure definitions.
\begin{lstlisting}
Lemma CCC_To_bproj_Sel_p : forall Defs C s C' s' p q l, str_projectable Defs C p ->
  CCC_To Defs C s (CCBase.TL.R_Sel p q l) C' s' ->
  exists Bp, bproj Defs C p = XSel q l Bp /\ bproj Defs C' p = Bp.

Lemma CCC_To_bproj_Call_p : forall Defs C s C' s' p X Xs, str_projectable Defs C p ->
  (forall Y, In Y Xs -> str_projectable Defs (snd (Defs Y)) p) ->
  (forall Y, set_incl_pid (CCC_pn (snd (Defs Y)) (fun X => fst (Defs X)))
                       (fst (Defs Y))) ->
  In X Xs ->  CCC_To Defs C s (CCBase.TL.R_Call X p) C' s' ->
  bproj Defs C p = XCall (X,p)
  /\ Xmore_branches (bproj Defs (snd (Defs X)) p) (bproj Defs C' p).
\end{lstlisting}
These lemmas are simple to prove by induction on \lstinline+C+.
The tricky part is getting the hypotheses strong enough that the thesis holds, and weak enough that any well-formed program will satisfy them throughout its entire execution.

From these results, it follows that projectability is preserved by choreography reductions.
This property is needed even to state the EPP Theorem, since we can only compute projections of projectable programs.
\begin{lstlisting}
Lemma CCC_To_projectable : forall P Xs ps,
  Program_WF Xs P -> well_ann P -> projectable Xs ps P ->
  (forall p, In p ps -> str_projectable (Procedures P) (Main P) p) ->
  (forall p, In p (CCC_pn (Main P) (Vars P)) -> In p ps) ->
  (forall p X, In X Xs -> In p (Vars P X) -> In p ps) ->
  forall s tl P' s', (P,s) --[tl]--> (P',s') -> projectable Xs ps P'.
\end{lstlisting}
Some of the hypotheses from the previous lemmas are encapsulated in the first two conditions:
well-formedness of \lstinline+P+, which ensures that any runtime term \lstinline+RT_Call X qs C+ in
\lstinline+Main P+ only includes processes in \lstinline+qs+ that are declared to be used by
\lstinline+X+ (this trivially holds if \lstinline+Main P+ is initial, and is preserved throughout
execution); and well-annotation of \lstinline+P+, i.e., the processes used in any
procedure are a subset of those it declares.\footnote{Equality is not necessary, and it would make
  this property harder to prove.}
The remaining hypotheses state, as before, that
\lstinline+Xs+ and \lstinline+ps+ include
all processes and procedures relevant for executing \lstinline+Main P+.

Similarly, we prove that strong projectability is preserved by transitions.

\paragraph{Completeness.}
In the literature, completeness of EPP is proven by induction on the derivation of the transition performed by the choreography $C$.
For each case, we look at how a transition for $C$ can be derived,
and show that the projection of $C$ can make the
same transition to a network with more branches than the projection of $C'$.
The proof is lengthy, but poses no surprises.
\begin{lstlisting}
Lemma EPP_Complete : forall P Xs ps, Program_WF Xs P -> well_ann P -> forall HP,
  (forall p, In p ps -> str_projectable (Procedures P) (Main P) p) ->
  (forall p, In p (CCC_pn (Main P) (Vars P)) -> In p ps) ->
  (forall p X, In X Xs -> In p (Vars P X) -> In p ps) ->
  forall s tl P' s', (P,s) --[tl]--> (P',s') ->
  exists N tl', (epp Xs ps P HP,s) --[tl']--> (N,s')
    /\ Procs N = Procs (epp Xs ps P HP) /\ forall H, Net N >> Net (epp Xs ps P' H).
\end{lstlisting}
By combining with the earlier results on pruning, we immediately obtain the generalisation for
multi-step transitions.

\paragraph{Soundness.}
Soundness is proven by case analysis on the transition made by the network, and
then by induction on the choreography inside each case.
For convenience, we split this proof in separate proofs, one for each transition.
We omit the statements of these lemmas, since they include a number of technical hypotheses (similar to those in e.g. \lstinline+SP_To_bproj_Com+, which is used in the proof of the case of communication, but more complex).
By contrast with completeness, all these lemmas are complex to prove: each case requires around
300 lines of Coq code.
The proofs have similar structure, but are still different enough that adapting them can not be done mechanically.

The last ingredient is a lemma of practical interest on procedure names: each process only uses ``its'' copy of the original procedure names.
This lemma is not only crucial in the proof of the next theorem, but also interesting in itself: it
shows that the set of procedure definitions can be fully distributed among the processes with no
duplications.
\begin{lstlisting}
Lemma SP_To_bproj_Call_name : forall Defs Defs' ps C HC s N' s' p X,
  SP_To Defs' (epp_C Defs ps C HC) s (R_Call X p) N' s' ->
  exists Y, X = (Y,p) /\ X_Free Y C.
\end{lstlisting}

All these ingredients are combined in the proof of soundness of EPP.
\begin{lstlisting}
Lemma EPP_Sound : forall P Xs ps, Program_WF Xs P -> well_ann P -> forall HP,
  (forall p, In p ps -> str_projectable (Procedures P) (Main P) p) ->
  (forall p, In p (CCC_pn (Main P) (Vars P)) -> In p ps) ->
  (forall p X, In X Xs -> In p (Vars P X) -> In p ps) ->
  forall s tl N' s', (epp Xs ps P HP,s) --[tl]-->(N',s') ->
  exists P' tl', (P,s) --[tl']--> (P',s') /\ forall H, Net N' >> Net (epp Xs ps P' H).
\end{lstlisting}

Generalising this result to multi-step transitions requires showing that pruning does not eliminate
possible transitions of a network.
This is in general not true, but it holds when the pruned network is the projection of a
choreography.

\begin{lstlisting}
Lemma SP_To_more_branches_N_epp : forall Defs N1 s N2 s' tl Defs' ps C HC,
  N1 >> epp_C Defs' ps C HC -> SP_To Defs N1 s tl N2 s' ->
  exists N2', SP_To Defs (epp_C Defs' ps C HC) s tl N2' s' /\ N2 >> N2'.
\end{lstlisting}

The formalisation of EPP and the proof of the EPP theorem consists of 13 definitions, 110 lemmas, and
4960 lines of Coq code.
The proof of the EPP Theorem and related lemmas make up for around 75\%\ of this size.

\section{Discussion and Conclusion}

We have successfully formalised a translation from a Turing-complete choreographic language into a
process calculus and proven its correctness in terms of an operational correspondence.
This formalisation showed that the proof techniques used in the literature are correct, and
identified only missing minor assumptions about runtime terms that trivially hold when these are
used as intended.

To the best of our knowledge, this is the first time such a correspondence has been formalised for a
full-fledged (Turing-complete) choreographic language.
Comparable work includes a preliminary presentation on a certified
compiler from choreographies to CakeML~\cite{GA18}, which however deals only with finite
behaviours~\cite{G20}.
In the related realm of multiparty session types (where choreographies do not include computation),
a similar correspondence result has also been developed independently~\cite{CFGY21}.

The complexity of the formalisation, combined with the similarities between several of the proofs,
means that future extensions would benefit from exploiting semi-automatic generation of proof
scripts.

Combining these results with those from~\cite{CMP21} would yield a proof that SP is also
Turing complete.
Unfortunately, the choreographies used in the proof of Turing completeness in~\cite{CMP21} are not
projectable, but they can be made so automatically, by means of an amendment (repair)
procedure~\cite{CM20}.
In future work, we plan to formalise amendment in order to obtain this result.
        
\bibliographystyle{splncs04}
\bibliography{biblio}
\end{document}